\documentclass[iop]{emulateapj}

\usepackage{amsmath}
\usepackage{epsfig}
\usepackage{graphicx}
\usepackage{apjfonts}
\usepackage[colorlinks=true,citecolor=blue]{hyperref}
\newcommand{\vsini}{V\!\sin\,i}
\newcommand{\kms}{\,{\rm km\,s}^{-1}}
\newcommand{\etal}{et al.}
\newcommand{\loga}{\log({\rm age})}
\newcommand{\msun}{M_{\sun}}

\slugcomment{}

\shorttitle{TIME-DEPENDENT NONEXTENSIVITY AND STELLAR ROTATIONAL EVOLUTION}
\shortauthors{SILVA \etal}

\begin{document}

\title{TIME-DEPENDENT NONEXTENSIVITY ARISING FROM THE ROTATIONAL EVOLUTION OF SOLAR-TYPE STARS}

\author{Jos\'{e} R. P. Silva\altaffilmark{1}, Mackson M. F. Nepomuceno\altaffilmark{1},
Br\'{a}ulio B. Soares\altaffilmark{1} \& Daniel B. de Freitas\altaffilmark{2}}

\altaffiltext{1}{Departamento de F\'{i}sica, Universidade do Estado do Rio Grande do Norte, Mossor\'o--RN,
Brazil; {joseronaldo@uern.br}}

\altaffiltext{2}{Departamento de F\'{i}sica, Universidade Federal do Rio Grande do Norte, Natal--RN, Brazil}

\submitted{Received 2013 April 29; accepted 2013 August 27}

\begin{abstract}

Nonextensive formalism is a generalization of the Boltzmann--Gibbs statistics. In this formalism the
entropic index $q$ is a quantity characterizing the degree of nonextensivity, and is interpreted
as a parameter of long-memory or long-range interactions between the components of the system. Since its
proposition in 1988, this formalism has been applied to investigate a wide variety of natural
phenomena. In stellar astrophysics, theoretical distribution function based on nonextensive formalism
($q$--distributions) has been successfully applied to reproduce the distribution of stellar radial and
rotational velocity data. In this paper, we investigate the time variation of the entropic
index $q$ obtained from the distribution of rotation, $\vsini$, for a sample of 254 rotational data for
solar-type stars from 11 open clusters aged between 35.5\,Myr and 2.6\,Gyr. As a result, we have
found an anti-correlation between the entropic index $q$ and the age of clusters, and that the distribution of
rotation $\vsini$ for these stars becomes extensive for an age greater than about 170\,Myr. Assuming that
the parameter $q$ is associated with long-memory effects, we suggest that the memory of the initial angular
momentum of solar-type stars can be scaled by the entropic index $q$. We also propose a physical link
between the parameter $q$ and the magnetic braking of stellar rotation.

\end{abstract}


\section{INTRODUCTION}
\label{intro}

Open clusters are formed by stars at nearly the same distance, with the nearly the same chemical, and created
at essentially the same time. They are highly important in astrophysics because they serve as ``laboratories" to
test theoretical models related to the effects and characteristics of the stellar composition, evolution, and
environment. It is widely accepted that star clusters are formed from the fragmentation of large, dense and
rotating molecular clouds, and in this process the original cloud angular momentum is partially
transferred to the newly formed stars \citep{Ballesteros11a, Ballesteros11b, Tomisaka00}.

The forming stars experience the action of several mechanisms changing their angular momentum. These mechanisms
include spin-up due to contraction of the cloud, redistribution of spin momentum of the contracting cloud into
orbital momentum of components by fragmentation, turbulent transport of the kinetic momentum in differently
rotating clouds, and remotion of angular momentum by magnetohydrodynamical effects such as magnetic braking
and driving of outflows and jets \citep{Konigl91, Shuetal94, Siess99, Masahiro07}. All these mechanisms acting
in certain stages, or even during the whole stellar formation process establish the angular momentum with
which the star reaches the main sequence.

After the star formation, closer encounters including tidal forces owing to the Galactic
gravitational field, encounters with field stars, and interstellar clouds crossing the way of the star cluster
are expected to occur. Furthermore, structural changes in the momentum of inertia due to a likely rotational
decoupling between the core and the outer shells of the star, and loss of angular momentum to the stellar
magnetized winds can play an important role in the evolution of the stellar angular
momentum \citep[e.g.,][]{EndSof78, Kawaler88, Bouvier08, Scholz09, Spada11}. All these physical mechanisms
contribute to the gradual loss of memory of initial stellar angular momentum, and makes the determination of
the distribution of the stellar rotation a difficult task.

In fact, over the last six decades, many researchers have proposed different laws controlling the rotational
velocity distributions \citep[see][]{Brown50, Chandrasekhar50, Deutsch65, Deutsch70, Gaige93, Fukuda82,
Guthrie82}. Usually, the frequency function for the rotational velocity, $\vsini$, is assumed to be a
Maxwell--Boltzmann distribution law \citep[e.g.,][]{Deutsch65, Deutsch70}, which considers that the system has
no memory of past experience (the ergodic hypothesis). \citet{Soaresetal06} have modified the
basic hypothesis of statistical independence between the distributions associated with the stellar angular
momentum components and have proposed the nonextensivity of the stellar angular momentum \citep[see also the
discussion in][]{Carvalhoetal08}. In this context, the distribution function for $y\equiv\vsini$ is given by
 \begin{equation}
   \varphi_{q}(y)=y\left[1-(1-q)\frac{y^{2}}{\sigma^{2}}\right]^{1/(1-q)},
   \label{fqobv}
 \end{equation}
where $q$ is an entropic index that measures the nonextensivity of the system (see Section \ref{nonextform}),
and $\sigma$ is a mass-dependent parameter associated with the characteristic width of the distribution
\citep[see][]{Soares11}. At the limit $q\rightarrow1$, Equation (\ref{fqobv}) reproduces the
Maxwellian function proposed by \citet{Deutsch65}. This distribution function has been tested and has been
successful in describing the distribution of observed stellar rotation data \citep{Soaresetal06,
Carvalhoetal08, Carvalhoetal10, Santoro10, Soares11}. More recently, \citet{deFreitas13} have analyzed the
relationship between the rotation and star ages in the context of the Tsallis formalism, and have shown  that
relationship can be well reproduced using the nonextensive theory approach.

In this paper we report the results of the analysis of the relationship between the index $q$ and the
stellar age based on a sample of 254 rotational velocity $\vsini$ data for solar-type stars in open clusters.
In the next section, we summarize the basic idea of nonextensive formalism. The $\vsini$ data are outlined
in Section \ref{sample}. In Section \ref{theparameters}, we present the method to obtain the parameters $q$
and $\sigma$ from the $\vsini$ distribution, followed by the discussion of the relationship between $q$
and cluster ages in Section \ref{rel_q_age}. Finally, we summarize our main results in Section
\ref{conclusions}.

\section{NONEXTENSIVE FORMALISM}
\label{nonextform}

Nonextensive formalism was proposed in \citet{Tsallis88} and it has since been shown to be
useful for studying systems out of equilibrium with long-ranged interactions, long-ranged memories, or
which evolve in fractal-like space-time. In this formalism the entropy is defined by
\begin{equation}
\label{gen_ent}
 S_q\equiv\frac{k}{q-1}\left(1-\sum\limits_{i=1}^W\,p_i^q\right)\quad\textrm{with}\,q\in\Re,
\end{equation}
where $k$ is a positive constant, $p_i$ denotes the probability for occupation of $i$-th state,
and $W$ is the total number of the configurations of the system. As $q\rightarrow 1$, the nonextensive entropy
definition recovers the standard form
\begin{equation}
 S_{q\rightarrow1} = -k\sum\limits_i\,p_i \ln{p_i},
 \label{bolt_ent}
\end{equation}
where $k$ is the usual Boltzmann constant. This means that the definition of nonextensive entropy contains
Boltzmann--Gibbs statistics as a special case.

The entropic index $q$ value is determined by the microscopic dynamics of the system and characterizes the
degree of nonextensivity, according to the following pseudo-additivity rule for two independent systems $A$
and $B$:
\begin{equation}
 S_q(A+B)=S_q(A)+S_q(B)+\frac{1-q}{k}\,S_q(A)\,S_q(B),
 \label{ps_add}
\end{equation}
where $q>1$ and $q<1$ correspond to superextensivity and subextensivity, respectively. In the limit
$q\rightarrow 1$, it recovers the Boltzmann--Gibbs entropy, which is extensive.

In nonextensive formalism the parameter $q$ is interpreted as a parameter of long-memory or long-range
interactions between the components of the system, and larger values for $q$ emphasize these long-memory or
long-range interactions. Apart from the stellar rotation, nonextensive formalism has been successfully
applied to investigate a wide variety of natural phenomena, such as self-gravitating polytropic systems
\citep{Plastino93}, Fokker--Planck systems \citep{Stariolo93}, high energy collisions \citep{WilkWlod00},
models for earthquakes \citep{Silvaetal06}, and stellar radial velocity in open clusters
\citep{Carvalhoetal07}.

\section{THE SAMPLE}
\label{sample}

The work sample consists of 254 $\vsini$ data for solar-type stars from 11 Galactic open clusters
aged between 35.5\,Myr and 2.6\,Gyr. The rotational data have been selected from the survey of radial and
rotational velocities performed by \citet{Mermilliod09}. The measurements were made with the CORAVEL
spectrometer \citep{Baranne79,Benz81}, and the $\vsini$ values were obtained following the calibration of
\cite{Benz84}, which enables measurements of $\vsini$ for dwarf stars with an accuracy of about 1$\kms$. For
this study, we have selected only the stars that are considered cluster members, namely stars not
flagged as ``NM'' in the notes column of Table 11 of \citet{Mermilliod09}, and with a color index ranging in the
interval of $0.55\lesssim\bv\lesssim0.75$, which corresponds to the stellar masses ranging in the interval of 
$0.9\lesssim M/\msun\lesssim 1.1\msun$ \citep{Cox00}. We have also rejected the spectroscopic binary stars
because their rotational velocities can be affected by tidal effects
\citep[e.g.,][]{Zahn77}.

As we are analyzing the relationship between $q$ and the cluster ages, using a homogeneous sample of
cluster ages is fundamental in this analysis. The data from \citet{Kharchenko05} provides a uniform
scale of ages for all clusters we are studying since all their cluster ages were determined with
the same isochrone-based procedure. These authors have established that the accuracy of their cluster age sample is
around 0.2--0.25 in log(age) by considering the sources of uncertainty that potentially
affect their method. One of the main factors limiting the accuracy in determining their cluster
ages is that for many young clusters ($\lesssim 1$\,Gyr) the ages were estimated by fitting only one
cluster member star. Another study relatively more rigorous in determining the ages by isochrone fitting was
performed by \citet{Meynet93}. These authors used a homogeneous data set of more than 4000 stars with accurately
determined membership, distance, metallicity, binarity, and reddening to estimate the ages of 30 open clusters
between 4\,Myr and 9.5\,Gyr. The comparison between the ages estimated by \citet{Kharchenko05} and
\citet{Meynet93} for 28 clusters presents a good overall agreement, with a root mean square (rms) of 0.33 in
log(age). In the present study, we will use the homogeneous sample of cluster ages from \citet{Kharchenko05}.
However, as the accuracy of the cluster ages may significantly influence our conclusions, we will be more
careful about the possible range of individual cluster ages, and will adopt a range equal to twice the rms
value from the comparison with the \citet{Meynet93} data, instead of the accuracy established in
\citet{Kharchenko05}.

As the age of Hyades is not provided by those authors, due to a limitation in their technique, we have
considered the age of 787\,Myr provided by the database for stars in open clusters WEBDA database
(http://www.univie.ac.at/webda/). The $\bv$ color indices were also taken from WEBDA. The original survey in
\citet{Mermilliod09} contains 13 open clusters; however, two of them could not be analyzed in this work because
they have few $\vsini$ data in the color index range we are considering. They are IC\,2391 and NGC\,7092 with
only one and two $\vsini$ data, respectively. In Table 1 we present the main characteristics of our work sample.
The number of stars in each cluster whose rotations are analyzed is given in the Column (2), the mean color
index and the mean rotational velocity in $\kms$ for these stars are shown in Columns (3),
and (4), respectively, and Column (5) presents the logarithm of the cluster age in years.

\begin{deluxetable*}{r|rrrr|rrrrr|c}
\tabletypesize{\scriptsize}
\tablecaption{Open Clusters with Rotational Velocity Analyzed in This Work, Best-fit Parameters, and Some
Other Information.\label{tab1}}
\tablewidth{0pt}
\tablehead{
\multicolumn{1}{c|}{(1)} &\multicolumn{1}{c}{(2)} & \multicolumn{1}{c}{(3)} &
\multicolumn{1}{c}{(4)} & \multicolumn{1}{c|}{(5)} & \multicolumn{1}{c}{(6)} &
\multicolumn{1}{c}{(7)} & \multicolumn{1}{c}{(8)}& \multicolumn{1}{c}{(9)} & \multicolumn{1}{c|}{(10)}&
\multicolumn{1}{c}{(11)}\\
 \multicolumn{1}{r|}{Cluster} & \multicolumn{1}{c}{$N$} & \multicolumn{1}{c}{$\bv$\tablenotemark{a}} &
\multicolumn{1}{c}{$\vsini$\tablenotemark{a}} & \multicolumn{1}{c|}{log(age)} &
 \multicolumn{1}{c}{$q$} & \multicolumn{1}{c}{$q/q_b$} & \multicolumn{1}{c}{$\sigma$} &
\multicolumn{1}{c}{$\sigma/\sigma_b$} & \multicolumn{1}{c|}{RSS\tablenotemark{b}} &
\multicolumn{1}{c}{Prob.}
}
\startdata
AlphaPer  & 14 & 0.64 $\pm$ 0.07 & 24.9 $\pm$ 17.6 & 7.55 & 1.460 $\pm$ 0.072 & 1.000 & 14.121 $\pm$ 1.861 &
1.001 & 0.015 & 0.678 \\
NGC\,1976 & 8  & 0.63 $\pm$ 0.05 & 12.6 $\pm$ 11.1 & 7.71 & 1.508 $\pm$ 0.079 & 0.999 &  5.132 $\pm$ 0.789
& 1.001 & 0.007 & 0.671 \\
IC\,2602  & 11 & 0.61 $\pm$ 0.07 & 19.6 $\pm$ 17.2 & 7.83 & 1.462 $\pm$ 0.158 & 0.995 & 9.841 $\pm$ 2.595 &
1.004 & 0.024 & 0.634 \\[1.5ex]
Pleiades  & 41 & 0.65 $\pm$ 0.07 & 15.9 $\pm$ 12.3 & 8.08 & 1.185 $\pm$ 0.085 & 0.997 & 13.215 $\pm$ 1.095 &
1.003 & 0.017 & 0.628 \\
NGC\,6475 & 13 & 0.68 $\pm$ 0.08 & 15.8 $\pm$  9.4 & 8.22 & 1.188 $\pm$ 0.108 & 1.001 & 13.868 $\pm$ 1.524 &
1.002 & 0.007 & 0.646 \\
Blanco 1  & 26 & 0.64 $\pm$ 0.07 & 12.3 $\pm$ 10.1 & 8.32 & 1.015 $\pm$ 0.070 & 0.996 & 12.067 $\pm$ 0.670 &
1.001 & 0.006 & 0.685 \\[1.5ex]
ComaBer   & 8  & 0.61 $\pm$ 0.07 &  9.4 $\pm$  5.7 & 8.78 & 1.078 $\pm$ 0.181 & 0.997 & 8.341 $\pm$ 1.222 &
1.001 & 0.003 & 0.671 \\
Hyades    & 43 & 0.63 $\pm$ 0.08 &  7.5 $\pm$  5.7 & 8.90 & 1.081 $\pm$ 0.109 & 0.998 & 6.878 $\pm$ 0.615 &
1.002 & 0.033 & 0.600 \\
Praesepe  & 43 & 0.65 $\pm$ 0.07 &  6.7 $\pm$  3.1 & 8.90 & 0.759 $\pm$ 0.095 & 0.988 & 8.646 $\pm$ 0.475 &
1.005 & 0.097 & 0.516 \\[1.5ex]
NGC\,0752 & 19 & 0.59 $\pm$ 0.04 &  5.7 $\pm$  3.3 & 9.14 & 1.015 $\pm$ 0.157 & 1.001 & 6.147 $\pm$ 0.754 &
1.000 & 0.015 & 0.657 \\
NGC\,2682 & 27 & 0.61 $\pm$ 0.06 &  5.3 $\pm$  2.8 & 9.41 & 0.772 $\pm$ 0.214 & 0.973 & 6.975 $\pm$ 0.869 &
1.010 & 0.111 & 0.618
\enddata
\tablenotetext{a}{Mean value for stars in the cluster $\pm$ standard deviations from the mean value.}
\tablenotetext{b}{The residual sum of squares measuring the discrepancy between the data and the
estimation model.}
\end{deluxetable*}

\section{PARAMETERS $q$ AND $\sigma$ FROM THE $\vsini$ DATA}
\label{theparameters}

The Empirical Cumulative Distribution Function (ECDF) for each cluster was computed as follows: (1) First, we sorted the
$\vsini$ data set in ascending order, from the smallest $\vsini_{a=1}$ to the largest $\vsini_{a=n}$, where
$n$ is the amount of $\vsini$ data for the cluster. (2) Then we computed the unbiased parameter $f_{n}(y)$ defined as
\begin{align*}
 f_{n}(y) &=
  \begin{cases}
   0        & \text{if } y \leq y_1 \\
   a/n      & \text{if } y_a < y \leq y_{a+1}, \\
   1        & \text{if } y > y_n
  \end{cases}
\end{align*}
where $y\equiv\vsini$. (3) Finally, the ECDF was made by plotting the computed parameter $f_{n}(y)$ versus $y$ values \citep{Qin98}.

To determine the best-fit parameters $q$ and $\sigma$, we first found a non-linear relation between
the variables $f_{n}$ and $\vsini$ by computing a Nadaraya--Watson kernel regression estimator (KDE) from the
ECDF with a Gaussian kernel and an appropriate bandwidth selected according to Scott's rule
\citep{Nadaraya64, Watson64, Scott92}. Next, we determined the best-fit parameters $q$ and $\sigma$ by
fitting the integral of Equation (\ref{fqobv}) to the ECDF. To adjust the curve, we used a Gauss--Newton
algorithm to minimize the residual sum of squares and a convergence tolerance of $10^{-5}$ \citep{Bates92}. In
order to reduce the risk of convergence to a local minimum, we used a graphical display to approximately
match the curve to the KDE by manually adjusting the parameters $q$ and $\sigma$. The values that produce the
visually matched curve were then used as the starting values. In addition, we tested the convergence, by making
the best-fit parameters $q$ and $\sigma$ vary randomly until 30\% from its original value and using these new
random values to start the procedure of fit. For all clusters, this process was repeated 300 times, and the
parameters always converged to the original best-fit values. The best-fit curves for each ECDF are presented
in Figure \ref{ecdfs}.

In order to constrain the best-fit parameters $q$ and $\sigma$ to a 95\% confidence interval, we used
bootstrap resampling with 1000 replicates to estimate the errors. The best-fit parameters,
$q$ and $\sigma$, and their error values are presented in Columns (6) and (8) of Table 1. In Columns
(7) and (9), we compare the best-fit values for $q$ and $\sigma$ with the same parameters from bootstrap
resampling, $q_b$ and $\sigma_b$, respectively. Finally, as the KDE provides an independent probability
estimation for each ECDF, we compared the best-fit curve with the KDE. For each cluster, we have
performed an Anderson--Darling $K$--sample test adjusted for ties \citep{ScholzSte87} with a 95\% confidence
level, and the null hypothesis that the best-fit curve and the KDE were drawn from the same probability
distribution. The null hypothesis was accepted for all clusters, according to the probability values shown in
Column (11) of Table 1.

\subsection{Low $\vsini$ Values}

The values of $\vsini$ from \citet{Mermilliod09} were determined by using the calibration of
\citet{Benz84}. Such a calibration allows the separation of the stellar rotation from the intrinsic sources
of line broadening such as micro-- and macroturbulence, pressure, and Zeeman splitting, and enables the
measurement of $\vsini$ for dwarf stars with an accuracy of about 1$\kms$, even for rotators with
$\vsini\approx2\kms$ \citep[c.f.][]{Benz84}. However, as reported by \citet{Quelozetal98}, a precise
separation of the rotation from all other broadening mechanisms affecting the stellar spectral lines is
crucial for determining the $\vsini$ values of rotators below $\sim 5\kms$. In this work, the
overestimation of $\vsini$ for low rotators will cause narrowing of the ECDF, which results in a systematic shift of the
best-fit $q$ to lower values. In order to evaluate the influence of this possible systematic error in
determining the rotation of slow rotators in our results, we proceeded as follows. We have performed for each cluster an
Anderson--Darling $K$--sample test with a 95\% confidence level to compare a sample such that all data with rotation of 
less than $5\kms$ were excluded with the original sample, containing all the data of rotation of the cluster.
The test results have showed that except for Hyades and NGC\,2682, the two samples for each cluster come from
a common population. For Hyades and NGC\,2682, we determined the $q$ values by considering all the slow
rotators ($\vsini\textless 5\kms$) as non-rotators ($\vsini = 0\kms$). Using the procedures described in Section
\ref{theparameters}, we obtained the best fit values ($q= 1.296\pm 0.162$; $\sigma=5.351\pm 1.07$),
and ($q=0.8042\pm 0.303$; $\sigma=6.789\pm 1.646$) for Hyades and NGC\,2682, respectively. These values
are very similar to these ones presented in Table \ref{tab1}, when we take into account the estimated errors. Then, possible
systematic errors in the determination of $\vsini\lesssim 5\kms$ do not seem to significantly affect our
results.

\section{THE RELATIONSHIP BETWEEN $q$ AND STELLAR AGES}
\label{rel_q_age}

\begin{figure}[h]
 \epsscale{1.0}
 \plotone{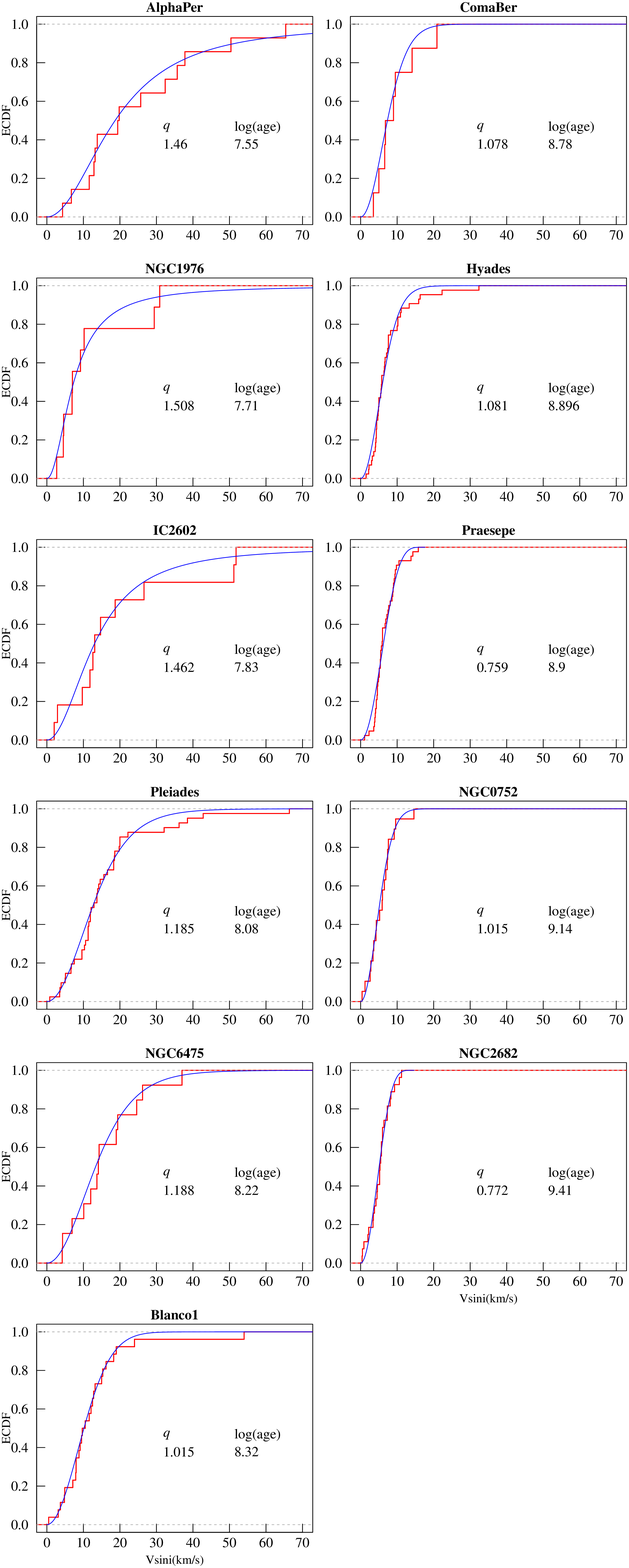}
 \caption{Empirical cumulative distribution functions for open clusters (red line) and their
best-fit curves (blue line). \label{ecdfs}}
 \end{figure}

\begin{figure}[ht]
 \epsscale{1.0}
 \plotone{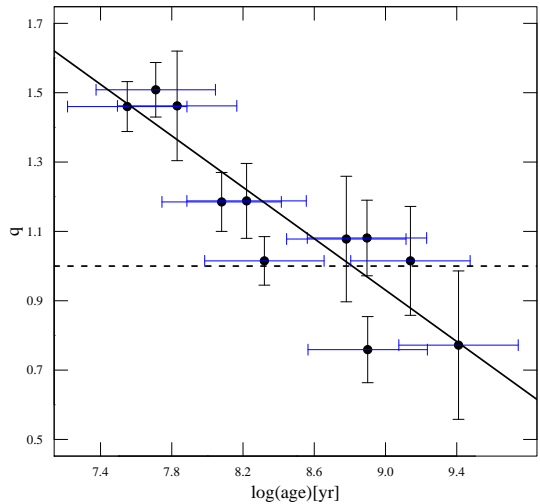}
 \caption{Vertical error bars represent the 95\% confidence interval from bootstrap resampling with 1000
replicates, and the horizontal ones are the rms = 0.33\,dex from the comparison between the ages
estimated by \citet{Kharchenko05} and \citet{Meynet93}. The continuous and dashed lines are the best-fit
curve $q=-0.37\loga +4.27$, and $q=1$, respectively. The point where these two lines intersect
corresponds to $\loga=8.84$\,dex (692\,Myr). \label{qxages}}
 \end{figure}

Figure \ref{ecdfs} displays the ECDF of $\vsini$ (dashed line) for each cluster, and the respective best-fit
curves (solid line). The changes to the shape of the ECDF with cluster age is phenomenologically well
understood. The younger clusters have a population of high rotators, and they present a higher $\vsini$
mean, and a longer tail to high velocities. As the cluster ages, this population evolves to slow rotators
\citep{Skumanich72, Kawaler88, Barnes03, IrwBouv09}, the $\vsini$ mean decreases, and the ECDF becomes
steeper. The $q$ values are correlated with the mean $\vsini$ values. We used a Spearman's $\rho$
statistic to estimate a rank-based measure of correlation between $q$ and mean $\vsini$. Such a test gives
the measure of correlation between variables as the parameter $\rho$ ranging in the interval $[-1,
1]$, with $\rho= 0$ indicating no correlation \citep{Hollander73}. The result gives $\rho=0.81$ with a
probability $p=0.001$ for the null hypothesis that the true $\rho$ is lesser than 0. As the index $q$ is
sensitive to the mean $\vsini$ of the distribution it is also sensitive to the cluster ages. The
time-variation of the nonextensivity of the $\vsini$ distributions as measured by $q$ emerges most clearly
from Figure \ref{qxages}, where we have plotted the index $q$ as a function of the cluster ages. There is an
anti-correlation between these two parameters. The Spearman's $\rho$ statistic gives $\rho = -0.88$, and $p=0.0002$
for the null hypothesis that the true $\rho$ is greater than 0.

\begin{figure}[ht]
 \epsscale{1.0}
 \plotone{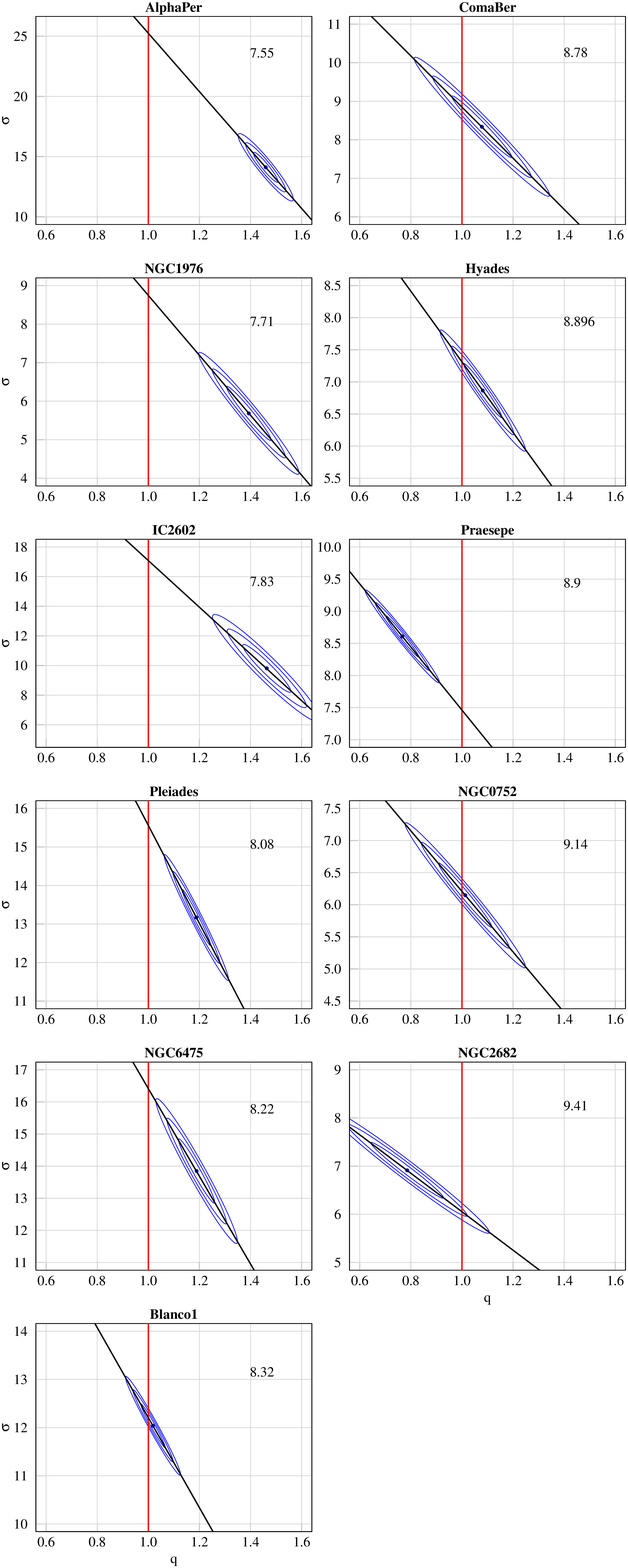}
 \caption{The 68.3\%, 95.5\% and 99.7\% confidence ellipses in the $q-\sigma$ plane. The red and black lines show
$q=1$ and the best-fit curve for the distribution of $\sigma$ as a function of $q$ values, respectively.\label{ellipses}}
 \end{figure}

The parameter $\sigma$ is associated with the characteristic width of the $\vsini$ distribution, and
therefore we expect that it is also correlated with the cluster ages. Spearman's $\rho$
statistic returns a correlation between $\sigma$ and cluster ages of $\rho = -0.46$, and a probability
$p=0.08$ for the null hypothesis that the true $\rho$ is greater than 0. In order to analyze whether this
correlation influences our results, we have drawn the 68.3\%, 95.5\%, and 99.7\% confidence ellipses on the
$q-\sigma$ plane, as shown in Figure \ref{ellipses}. The confidence ellipses were computed for 1000
bootstrap replications of $q$ and $\sigma$ parameters for each cluster. According to Figure
\ref{ellipses}, we can state with a confidence level of 99.7\% that the nonextensivity of the $\vsini$
distributions ($q > 1$) is present until about the age of NGC\,6475: $\loga=8.22$\,dex $\sim 170$\,Myr.
If we consider the age range adopted for Blanco 1, our sample will give a minimum age limit to
nonextensivity of $\loga=7.99$\,dex $\sim 100$\,Myr.

The nonextensivity (i.e., $q\neq1$) of the distribution of stellar rotation, $\vsini$, was reported by
\citet{Soaresetal06, Carvalhoetal08, Carvalhoetal10, Santoro10}; and \citet{Soares11}. Now
we state that for solar-type stars in open clusters: 1) the nonextensivity of the distribution of
$\vsini$ decreases when the cluster ages, and 2) the distribution of $\vsini$ becomes extensive for ages of
about 700\,Myr. As mentioned in Section \ref{nonextform}, the entropic index $q$ measures the degree of
nonextensivity of the system and emphasizes effects associated with long-memory or long-range interactions. In this context,
one possible explanation for the anti-correlation between $q$ and cluster age would be the memory of the
initial angular momentum of the stars vanishing and thus, this memory loss would be scaled by the entropic
index $q$. We recall that the solar-type stars retain the memory of their initial angular momentum until an
age of greater than 100 Myr, with this age being even higher for lower mass
stars \citep{Pinso90, Bouvier97, Scholz07, Bouvier09, Scholz09, IrwBouv09}. We estimate that the age for
extensivity, when stars have lost all memory of past angular momentum history, is around
700\,Myr considering the best-fit curve presented in Figure \ref{qxages}.

In their study, \citet{deFreitas13} have shown that the rotation--age relationship can be reproduced using a
model from nonextensive formalism. Their model, based on the nonextensivity parameter $q_K$, derived
from the Kawaler's parameterization, has been tested using $\vsini$ data for F- and G- type
main-sequence field stars with ages limited to 10\,Gyr, and has closely reproduced the decrease of rotation
with age. These authors have proposed that the parameter $q_K$ is associated with the dynamo process and
magnetic field geometry. In their model, $q_K=1$ indicates that the stars are in a saturated magnetic field
regime, while higher $q_K$ values denote an unsaturated domain, where magnetic braking law can be applied.
In this context, by considering the correlation between $q$ and the mean $\vsini$ values, Figure \ref{qxages}
could indicate that cluster stars in our sample are in an unsaturated magnetic field domain ($q_K > 1$),
where the effects of magnetic torque are present.

\begin{figure}[htp]
 \epsscale{1.0}
 \plotone{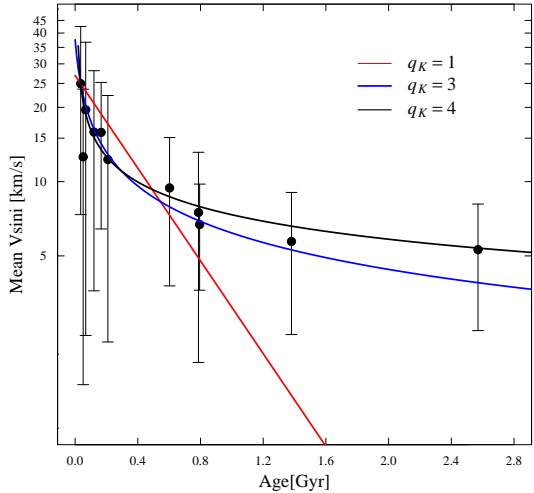}
 \caption{Mean $\vsini$ as a function of the cluster ages. The black line represents the best-fit for the model
proposed by \citet{deFreitas13}. The red and blue lines show the best-fit curves for an exponential decay and the
Skumanich law, respectively. \label{qkxages}}
 \end{figure}

In Figure \ref{qkxages}, we present the distribution of mean $\vsini$ as a function of age as well as the
curve obtained by fitting the model proposed by \citet{deFreitas13} to the data (black line). The
best-fit values for $q_K$ and the nonextensive Lyapunov exponent coefficient are $q_{K}=4.06\pm0.78$, and
$\lambda=13.95\pm6.27$, respectively, with a log-likelihood value $LogLik=-23.46$. In addition, Figure
\ref{qkxages} displays the best-fit curves for $q_K=1$, $\lambda=2.16\pm0.55$, $LogLik=-31.51$ (red line), and
$q_K=3$, $\lambda=7.89\pm1.66$,  $LogLik=-23.47$ (blue line), representing an exponential decay, and
Skumanich law \citep{Skumanich72}, respectively. The unsaturated model with $q_K=4$ as well as Skumanich law
have smaller $LogLik$ magnitudes, indicating that these models are better than the exponential decay in
fitting the data.

According to Kawaler's parameterization \citep{Kawaler88}, $a$ and $N$ are parameters
related to the dynamo theory and topology of the magnetic field, respectively. The $a$--parameter ranges
between 1 and 2, for the unsaturated domain, and is 0, for the saturated one. The saturated regime is
described by a usual exponential ($q_{K}=1$ in the nonextensive approach). The model proposed by
\citet{deFreitas13} suggests that the unsaturated regime is driven by a power law with exponent
$q_{K}=1+4aN/3$. For this regime, $q_{K}$ presents a lower value denoted by $q_{K}=1+4N/3$, when $a=1$, and an
upper value $q_{K}=1+8N/3$, when $a=2$. Then, for $q_{K}\sim 4$, as obtained from our sample, the topology of
magnetic field is essentially radial, with $N$ ranging from $1.1$ to $2.3$. For Skumanich law, namely
$q_{K}=3$, depending on the value of the parameter $a$, the topology of the magnetic field can be either radial,
with $N=1.5$, or dipolar, with $N=0.75$. These values for parameter $N$, are approximately equal to those
obtained by \citet[][see Table 1]{deFreitas13}, if we consider the same mass range of the stars in our sample.
In the context of the theory of magnetic braking, these results support the hypothesis that
solar-type stars are under the action of a braking torque provided by a radial or dipolar magnetic field.
Accordingly, recent numerical simulations for solar-like stellar winds from rotating stars with dipolar
magnetic fields carried out by \citet{matt2012} have shown that low or moderate
rotation rates do not influences significantly the net torque on the star since at this rotational regime the
the Alfv\'{e}n radius\footnote{\citet{matt2012} define the ``Alfv\'{e}n radius" as the cylindrical radial
location where the wind velocity equals the local Alfv\'{e}n speed.} is nearly independent of the spin rate.

In order to establish the relationship between the parameter $q_K$, derived from Kawaler's
parameterization, and the parameter $q$ from the distribution of $\vsini$, we suggest the equation $q\approx
q_0(1 -\Delta t/q_K)$. The variation $\Delta t=t-t_0$ is the difference between the logarithms of the cluster
ages, and $q_0$ is the index $q$ for the cluster with age $t_0$. This equation is a heuristic approximation based on the
observation that when $q_K\sim4$, it nearly matches the best-fit curve in Figure \ref{qxages}. There are at
least two important consequences in admitting this relationship between $q_K$, $q$, and cluster age. First,
the parameter $q_K$ will determine the slope of the curve $q$ versus cluster age. Second, we will have a
bridge between the index $q$ and the theory of magnetic braking of stellar rotation, since $q_K$ is related
to the parameters $a$ and $N$.

Finally, it is worth mentioning that in a forthcoming paper we will present the results of applying this
analysis to a large data set of rotational periods for solar mass stars in open clusters over the age range 
1\,Myr to $\sim 1$\,Gyr \citep[e.g.,][]{Herbst02, Meibom11}. The rotation periods have
multiple advantages over $\vsini$ as, for example, the former is free of the ambiguity inherent to the
$\sin\,i$, and since stellar rotational periods can be measured with high precision from spot-modulation,
the slower rotators are better represented in the database.

\section{CONCLUSIONS}
\label{conclusions}

We have analyzed the time relationship between the entropic index $q$ from the empirical distribution of
rotational velocities $\vsini$ for solar--type stars from 11 open clusters aged between 35.5\,Myr and
2.6\,Gyr. As a result, we have found that there is an anti--correlation between these $q$ values and the
cluster age. The entropic index $q$ in Tsallis formalism is a measure of the degree of nonextensivity of a
system, with larger values for $q$ emphasizing the long-memory or the long-range interactions. In this work we
have proposed that the loss of initial angular momentum memory for solar type stars in open clusters can be
scaled by the index $q$. In this context, our results indicate that these stars lose the memory of their
initial angular momentum for an age greater than about 170\,Myr, when their rotational distribution becomes extensive.
In addition, we have proposed a correlation between $q$ and the nonextensivity parameter $q_K$ from Kawaler's parameterization,
according to which $q_K$ determines the behavior of $q$ with cluster ages. Such a relation between these two
parameters can constitute a link between the parameter $q$ obtained from the distribution of $\vsini$
and the theory of magnetic braking of stellar rotation. Going forward, we believe that our results can
provide an additional support for using Tsallis nonextensive formalism to investigate the stellar angular
momentum evolution.

\acknowledgments
This study was partially funded by the Programa Institutos Nacionais de Ci\^encia e Tecnologia
(MCT-CNPq-Edital No. 015/2008). We should like to thank the anonymous referee for providing constructive
comments and suggestions that strengthened the manuscript and reported results.

\end{document}